\documentstyle[aps,psfig,multicol]{revtex}

\def\BM{\begin{multicols}{2}}
\def\EM{\end{multicols}}
\def\be{\begin{equation}}
\def\ee{\end{equation}}
\def\bea{\begin{eqnarray}}
\def\eea{\end{eqnarray}}
\newcommand{\PSFIG}[1]{\centerline{\psfig{file=#1}}}

\begin{document}

\title{Dynamics of a Limit Cycle Oscillator under Time Delayed 
Linear and Nonlinear Feedbacks}
\author{D. V. Ramana Reddy,\cite{DVRemail} A. Sen,\cite{ASemail} and 
G. L. Johnston\footnote{Present address:12 Billings St., Acton, 
MA 01720, USA.}\\
{\it Institute for Plasma Research, Bhat, Gandhinagar 382 428}}
\maketitle

\begin{abstract}
We study the effects of time delayed linear and nonlinear
feedbacks on the dynamics of a single Hopf bifurcation
oscillator. Our numerical and analytic investigations reveal a
host of complex temporal phenomena such as phase slips,
frequency suppression, multiple periodic states and chaos. Such
phenomena are frequently observed in the collective behavior of a
large number of coupled limit cycle oscillators. Our time
delayed feedback  model offers a simple paradigm for
obtaining and investigating these temporal states in a single 
oscillator.
We construct a detailed bifurcation diagram of the oscillator as
a function of the time delay parameter and the driving strengths
of the feedback terms. We find some new states in the presence
of the quadratic nonlinear feedback term with interesting
characteristics like birhythmicity, phase reversals, radial trapping,
phase jumps, and spiraling patterns in the amplitude space.
Our results may find useful applications in physical, chemical
or biological systems.


\end{abstract}

\BM
\section{Introduction}
Coupled limit cycle oscillators have been extensively studied in
recent times as a mathematical model for understanding the
collective behavior of a wide variety of physical, chemical and
biological 
problems \cite{Mcc:71,DL:73,Win:80,Win:87,KN:87,BSWSH:89,%
GHSS:92,DMDS:93,CS:93,Dai:96,Pec:98,NS:80}.
One of the simplest and earliest of such
models is the so called Kuramoto model \cite{KN:87}, which is a mean 
field model of a collection of phase oscillators, and clearly 
exhibits such cooperative phenomenon as spontaneous synchronization 
of the oscillators beyond a certain coupling strength. A more
generalized version of the coupled oscillator model that includes both
phase and amplitude variations exhibits collective behavior like 
amplitude death, where, for a large enough spread in the natural
frequencies of the oscillators, an increase in the coupling
strength induces a stabilization of the origin, leading to a
total cessation of oscillations in the 
system \cite{Bar:85,AEK:90,MS:90,MMS:91}.
Other collective
states observed in these models include partial
synchronization, phase trapping, large amplitude Hopf
oscillations and even chaotic behavior \cite{MS:90,MMS:91}.
Recently there has been
some interest in investigating the effect of time delay on the
collective dynamics of these coupled models
\cite{SW:89,NSK:91,NTM:94,KPR:97,RSJ:98,RSJ:99,YS:99,BC:99}. Time
delay is ubiquitous in most 
physical \cite{CHP:36,Erg:54,Dri:63,WWPOG:94},
chemical \cite{MY:99}, biological \cite{TKRGH:96},
neural \cite{Des:94}, ecological \cite{Cus:76}, and other natural
systems due to finite propagation speeds of signals, finite
processing times in synapses, and finite reaction times.  Time
delayed coupling introduces interesting new features in the
collective dynamics, e.g. simultaneous existence of several different
synchronized states \cite{SW:89,NSK:91,NTM:94,KPR:97}, 
regions of amplitude
death even among identical oscillators \cite{RSJ:98,RSJ:99},
and bistability between synchronized and incoherent
states \cite{KPR:97,YS:99}.

One of the remarkable aspects of this
cooperative dynamics is that many of its salient features can
be observed even in a system consisting of just two coupled
oscillators \cite{AEK:90,SW:89,RSJ:98,RSJ:99}. The
temporal behavior of either of the two oscillators in such a
case (which is easy to investigate both numerically and analytically) 
reveals a great deal about the collective aspects of larger systems.
In fact, a useful point of view to adopt is to regard each
oscillator as being driven autonomously by a source term that
represents the collective feedback of the rest of the system.
Motivated by such a qualitative consideration, we have studied
in detail the dynamics of the following model system of an 
autonomously driven single limit cycle oscillator
\be
\label{EQN:E1}
\dot{Z}(t) - (a + i \omega - \mid Z(t) \mid^2) Z(t) = f(Z(t-\tau)),
\ee
where $Z(t) = X + iY$ is a complex quantity, 
$\omega$ the frequency of oscillation, $a$ a real constant,
and $\tau \ge 0$ is the time delay of the autonomous feedback term
$f$.  
In the absence of the feedback term
Eq.~(\ref{EQN:E1}), often called the Stuart-Landau equation,
has a stable limit cycle of amplitude $\sqrt{a}$ with angular frequency
$\omega$. It is simply the normal form of a supercritical Hopf
bifurcation and is a useful nonlinear model for a variety of
physical, chemical and biological systems.
For the autonomous feedback term we choose the following
model form:
\be
\label{fbterm:E2}
f(Z(t-\tau)) =  - K_1 Z(t-\tau) - K_2 Z^2(t-\tau),
\ee
where $K_1$ and $K_2$ represent the strengths of the linear and
nonlinear contributions of the feedback. This choice is 
motivated by considerations of both mathematical simplicity
and possible importance for modeling of physical and biological systems.
The quadratic term is the simplest nonlinearity that can break the 
rotational symmetry of the Stuart-Landau system. Physically this term
introduces nonlinear mode coupling, a process that is important in 
large coupled systems.  
Eq. (\ref{EQN:E1}) can also be viewed as a prototype equation 
arising in the delayed feedback control of an individual
physical or biological entity that can be modeled by the normal
form. Our results may thus be of more general and direct 
utility in addition to provide useful insights into the collective 
dynamics of large systems. Similar studies (using a variety of
feedback terms) exist for the damped harmonic oscillator, e. g.
\cite{SC:95}, but we are not aware of such investigations
for our model limit cycle oscillator. A few investigations in
the past have restricted themselves to the study of noise and
perturbations \cite{FL:85,MLL:90,KS:91} on the dynamics of such
an oscillator.

The organization of our paper is as follows. In Section
\ref{sec:lin}, we analyze the dynamics of the oscillator using
just the linear feedback term and discuss the analytic
conditions for the stability of the origin and the existence of
periodic orbits. Detailed bifurcation diagrams are plotted as a
function of the various system parameters like $a$, $K$ and
$\tau$ and their similarity to collective states of larger
systems is pointed out. We also present  numerical results on
higher frequency states, which can coexist with the lowest
periodic state and discuss the phenomenon of frequency suppression of
these states as a function of the time delay parameter. Section
\ref{sec:nlin} treats the full feedback term by including the
quadratic nonlinear contribution. The bifurcation diagram is a
great deal richer now due to the existence of two other
equilibrium points in addition to the origin. We analyze the
stability of these equilibria and the consequent temporal
behavior of the oscillator in various parametric regimes. Some
novel temporal states are pointed out. Section \ref{sec:con}
summarizes our results and discusses their significance and
possible applications.

\section{Time delayed linear feedback}
\label{sec:lin}
We begin our analysis of the model 
Eq. (\ref{EQN:E1}) by considering only the
linear feedback term (i.e. $K_2 =0$), so that we have
\be
\label{EQN:E3}
\dot{Z}(t) = (a + i \omega - \mid Z(t) \mid^2) Z(t) - K Z(t-\tau),
\ee
where we have put $K_1=K$ for simplicity of notation.
Note that the above linear feedback term is similar in form to 
the feedback term used extensively in experimental and theoretical 
investigations of control of chaos using the 
Pyragas method \cite{Pyr:92}.
The actual form in the Pyragas method is $[Z(t-\tau) - Z(t)]$,
which is equivalent to replacing the constant $a$ by $a + K$ in the
above equation. However, unlike the systems investigated for 
the Pyragas method, Eq. (\ref{EQN:E3}) has no regimes 
of chaotic behavior. 
We will examine instead the effect of the time delayed feedback 
on the stability of the origin and on the nature of the 
periodic solutions.

In the absence of time delay, 
it is clear from inspection that Eq. (\ref{EQN:E3})
has a time-asymptotic periodic solution given by 
$Z(t) = \sqrt{a - K} e^{i \omega t}$ for $a>K$. If $a<K$, then the 
origin is the only stable solution; i.e., no 
oscillatory time-asymptotic solutions are possible. 
At $a=K$, the oscillator undergoes a 
supercritical Hopf bifurcation. We now analyze
systematically the effect of time delay on the
stability of the origin and the periodic solutions.

\subsection{Stability of the origin}
\label{sec:stab}
The origin $Z_p = (0, 0)$ is a fixed point of Eq. (\ref{EQN:E3}).
To study its stability we assume that the perturbations about $Z_p$
grow as $e^{\lambda t}$, where $\lambda$ is
a complex number. Substituting in Eq. (\ref{EQN:E3}) and 
linearizing about $Z = Z_p$, 
one easily obtains the following 
characteristic equation:
\be
\label{EQN:char}
\lambda = a \pm i \omega - K e^{- \lambda \tau},
\ee
where the $\pm$ sign arises from considering the complex conjugate
of Eq. (\ref{EQN:E3}). This ensures that we have the complete
set of eigenvalues. For $\tau = 0$, one obtains  $\lambda = a -
K \pm i ~\omega$.  The origin is stable in the region of parametric
space where Re($\lambda$) $< 0$, which occurs when $K > a$. So
the critical, or the marginal stability curve
is given in this case by $K = a$.  When $\tau \ne 0$, 
Eq. (\ref{EQN:char}) remains a transcendental equation with a
principal term $\lambda ~e^{\lambda \tau}$ and hence the
equation possesses an infinite number of complex solutions.  Let
these roots be ordered according to the magnitude of the real
parts: $\{\ldots, \lambda_n,
\ldots, \lambda_2,
\lambda_1, \lambda_0\}$, where Re$(\lambda_{m-1}) < $ Re$(\lambda_m)$.
The problem of finding the stability
criterion then reduces to that of finding the conditions on 
$K, \omega$ and $\tau$ such that
Re$(\lambda_j) < 0$, for all $j$. 
Let $\lambda = \alpha + i \beta$, where $\alpha$ and $\beta$ are 
real. By substituting this in Eq. (\ref{EQN:char}), we get
\bea
\label{EQN:chara}
&\alpha& = a - K ~e^{-\alpha \tau} ~\cos(\beta \tau) , \\
\label{EQN:charb}
&\beta& = \pm \omega + K ~e^{-\alpha \tau} ~\sin(\beta \tau) .
\eea
We can arrive at the following two equations for $\alpha$ 
and $\beta$ by squaring and adding the above two equations and 
by dividing the first equation by the second respectively:
\bea
\label{EQN:beta}
&\beta& \equiv \beta_\pm = \omega \pm \sqrt{K^2 e^{-2\alpha \tau} - 
(\alpha-a)^2} , \\
\label{EQN:alpha}
&\alpha& = a - (\beta - \omega) / \tan(\beta \tau),
\eea
where, in the above and from here after, we consider only one
set of curves by choosing $\beta = +\omega \pm \sqrt{..}$ . The
other set of curves arising due to $\beta = - \omega \pm
\sqrt{..}$ is implicit in the above since the eigenvalues always
occur in complex conjugate pairs.  From Eq. (\ref{EQN:beta}) we
see that $\beta$ is real only when $(\alpha - a)^2 ~e^{2 \alpha
\tau} \le K^2$. So for any finite value of $K$, the value of
$\alpha$ is bounded from above. 
\EM
\begin{figure}[t]
\PSFIG{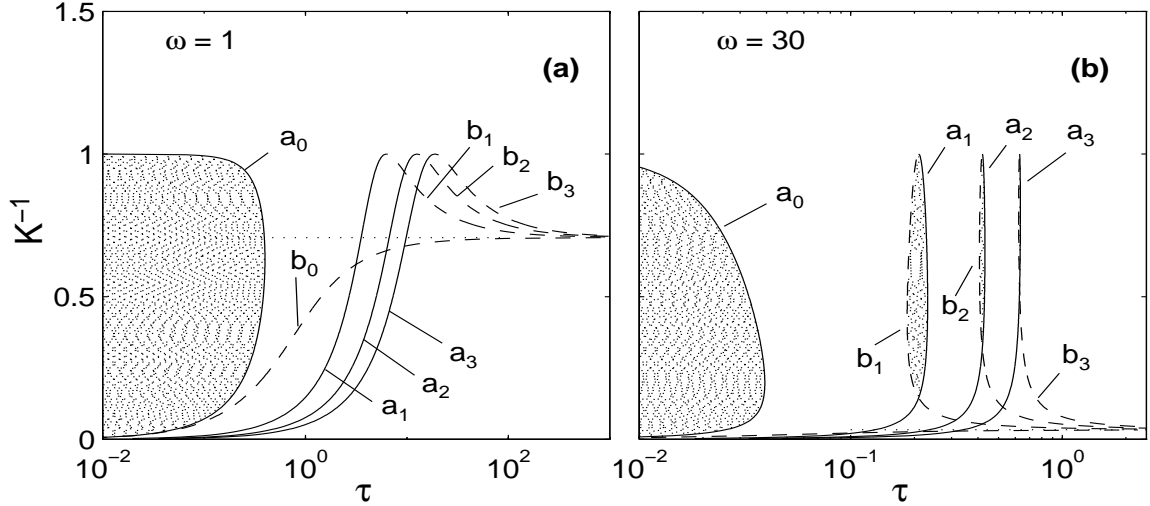,height=7cm,width=17cm}
\caption{%
The stability of the origin is shown in a $(\tau,K^{-1})$ diagram
for (a) $\omega = 1, a = 1$  and (b) $\omega = 30, a = 1$. 
At lower values of $\omega$ there is only one 
death region that is confined between $\tau = 0$ and 
$\tau = \tau_1(0,K)$, but for higher values of $\omega$ there 
are generally multiply connected regions of amplitude death.
The curves $a_n$ and $b_n$ represent $\tau_1(n,K)$ and 
$\tau_2(n,K)$ respectively. 
The dotted horizontal line is $1/f(\omega)$.
}
\label{FIG:death}
\end{figure}
\BM
To obtain the critical curves, set $\alpha = 0$. This gives
\be
\label{EQN:beta0}
\beta|_{\alpha = 0} = \omega \pm \sqrt{K^2 - a^2}.
\ee
By inverting Eq. (\ref{EQN:chara}), and 
noting that $\sin(\beta_+\tau) > 0$, $\sin(\beta_-\tau) < 0$, 
and that $\beta_-$ can be either positive or negative,
we obtain the following two sets of critical curves:
\bea
\label{EQN:tau1}
\tau_1(n,K) = \frac{2 n \pi + \cos^{-1}(a/K)} 
{\omega + \sqrt{K^2 - a^2}} , \\
\label{EQN:tau2}
\tau_2(n,K) = \frac{2 n \pi - \cos^{-1}(a/K)} 
{\omega - \sqrt{K^2 - a^2}} .
\eea
In Eq.~(\ref{EQN:tau1}), $n = 0, 1, 2, \ldots $. 
In Eq.~(\ref{EQN:tau2}), if $\beta_->0, ~n = 1, 2, \ldots $;
if $\beta_-<0, ~n = 0, 1, 2, \ldots $.
Thus the critical curves exist only in the
region $K \ge a$.  Since, for $\tau = 0$, the region of
stability of the origin is given by $K > a$, the corresponding
region, for $\tau > 0$, will be given by the area between $\tau
= 0$ and the critical curve closest to the line $\tau = 0$. This
critical curve should be the one on which $d\alpha / d\tau
> 0$. From (\ref{EQN:char})
\be
\frac{d\lambda}{d\tau} = \frac{K \lambda e^{-\lambda\tau}}
{1 - K \tau e^{-\lambda\tau}} ~~, 
\ee
and 
\bea
\frac{d\alpha}{d\tau}\Bigg|_{\alpha = 0} =
\mathrm{Re} \frac{K (i\beta) e^{-i\beta\tau}}
{1 - K \tau e^{-i\beta\tau}}
& = & \beta ~K \sin(\beta\tau) ~D^{-1} \nonumber \\
& = & \beta (\beta - \omega) ~D^{-1},
\eea
where $D = [1-K \tau \cos(\beta \tau)]^2 + 
[K \tau \sin(\beta \tau)]^2$ is 
positive real.  Hence
\be
{\frac{d\alpha}{d\tau}}\Bigg|_{\alpha = 0} 
\cases{ > 0 & on $\tau_1$, \cr
        > 0 & on $\tau_2$ if $K > f(\omega)$, \cr
        = 0 & on $\tau_2$ if $K = f(\omega)$, \cr
        < 0 & on $\tau_2$ if $K < f(\omega)$, \cr}
\ee
where $f(\omega) = \sqrt{a^2+\omega^2}$.
The above condition implies that there can be only one stability
region if $K > f(\omega)$.  There is a possibility of multiple
stability regions if $K < f(\omega)$. Our numerical plot
in Fig.~\ref{FIG:death}(a) of the
curves $\tau_1(n,K)$ and $\tau_2(n,K)$ reveals that the region
between $\tau = 0$ and $\tau = \tau_1(0,K)$ is the only
stability region possible for small values of $\omega$. However,
as the value of $\omega$ is increased, the stability regions can
be specified by $0 \le \tau < \tau_1(0,K)$ and $\tau_2(n,K) <
\tau < \tau_1(n,K)$, where $n > 0$. In Fig.~\ref{FIG:death}(b) 
the critical curves
are plotted from Eqs. (\ref{EQN:tau1}) and (\ref{EQN:tau2}) for such
a large value of $\omega$, namely $\omega = 30$, and the
multiple stability regions are represented by the shaded portions. 
Note that a similar situation arises in the case
of two or more limit cycle oscillators that are coupled
by a time delay. This was investigated in
detail in \cite{RSJ:98,RSJ:99}, where the collective 
stability regions were
termed {\it amplitude death} regions or {\it death islands} in
the $K -\tau$ space.

We conclude this section by carrying out a stability analysis of
the origin in the $(a,K)$ plane for a fixed value of $\tau$.
Using Eqs.(\ref{EQN:chara}-\ref{EQN:charb}) we write below the
critical curves which are non intersecting:
\be
K(\beta) = \frac{\beta \pm \omega}{\sin(\beta\tau)}, ~~and~~
a(\beta) = \frac{\beta \pm \omega}{\sin(\beta\tau)} ~\cos(\beta\tau).
\ee
We note that the above expressions for $K(\beta)$ and $a(\beta)$
have singularities at $\beta = n \pi/\tau$ and between any two 
successive singular points the expressions produce continuous curves 
in $(a,K)$ plane.  Following Diekmann et al. \cite{DVVW:95},
we define the following intervals where the sign of the superscript 
of $I$ indicates the sign of the function $\sin(\beta\tau)$ in 
that interval:
\be
I^{-}_{n} = \left((2n-1)\frac{\pi}{\tau}, ~2 n \frac{\pi}{\tau}\right),
~~~
I^{+}_{n} = \left(2n\frac{\pi}{\tau}, ~(2 n+1) \frac{\pi}{\tau}\right),
\ee
for $n = 0, 1, 2, \ldots .$ We restrict our attention, without loss of 
generality, to the case of $\beta \ge 0$.
Hence we can define the following curves in $(a,K)$ plane.
\be
C^{\pm}_n = \Bigg\{ (a,K) = 
\Bigg(\frac{(\beta-\omega)~\cos(\beta\tau)}{\sin(\beta\tau)}, 
~\frac{(\beta-\omega)}{\sin(\beta\tau)} \Bigg)
 ~\Bigg| ~\beta \in I^{\pm}_n \Bigg\} , 
\ee
\be
D^{\pm}_n = \Bigg\{ (a,K) = 
\Bigg(\frac{(\beta+\omega)~\cos(\beta\tau)}{\sin(\beta\tau)}, 
~\frac{(\beta+\omega)}{\sin(\beta\tau)} \Bigg) 
~\Bigg| ~\beta \in I^{\pm}_n \Bigg\} 
\ee
These curves are 
parametrized by $\beta$. The curves are degenerate
at $\omega = n \pi/\tau$. For $\omega = 2 n \pi/\tau, 
~~n = 0, 1, 2, \ldots $, both the curves
$C_n^\pm$ and $D_n^\pm$ merge, and there is another curve in 
addition to the above, defined by
\be
C_R = \{ (K,a) ~\mid~ a = K \}
\ee
with $\beta = 2 n \pi/\tau $. Likewise
for $\omega = (2 n + 1)\pi/\tau, ~~n = 0, 1, 2, \ldots $, the 
corresponding additional curve is defined by
\be
C_R = \{ (K,a) ~\mid~ a = - K \}
\ee
with $\beta = (2 n +1) \pi/\tau $. 
We are now only left with the task of finding out the number of
the eigenvalues in the right half plane on either side of the
curves $C_n^\pm$ and $D_n^\pm$.  For our particular problem 
it is possible to carry out this analysis in an exact manner. Let
$F(a,K,\lambda) = \lambda - a \mp i \omega + K e ^{-\lambda
\tau} = 0$ be the eigenvalue equation. Define $G_1 = $Re$F$, 
$G_2 =$ Im $F$, and at $\alpha = 0$ define a matrix $M$ by\\
\be
M = \left( \matrix{ 
\frac{\partial G_1}{\partial a} & 
\frac{\partial G_1}{\partial K} \cr \cr
\frac{\partial G_2}{\partial a} & 
\frac{\partial G_2}{\partial K} } \right) .
\ee
We now make use of a proposition of Diekmann et al. \cite{DVVW:95}
to determine the  positions of the eigenvalues in the complex
plane with respect to the curves $(a(\beta), K(\beta))$.
The proposition states that {\it the critical roots are in the
right half-plane in the parameter region to the left of the
curve $(a(\beta), K(\beta))$, when we follow this curve in the
direction of increasing $\beta$, whenever $\det M < 0$ and to
the right when $\det M > 0$.} In the present case, on both the
curves $C^\pm_n$ and $D^\pm_n$, the matrix is given by\\
\be
M = \left( \matrix{ -1 & \cos(\beta\tau) \cr  \cr
           0 & -\sin(\beta\tau)} \right) , 
\ee
\begin{figure}
\PSFIG{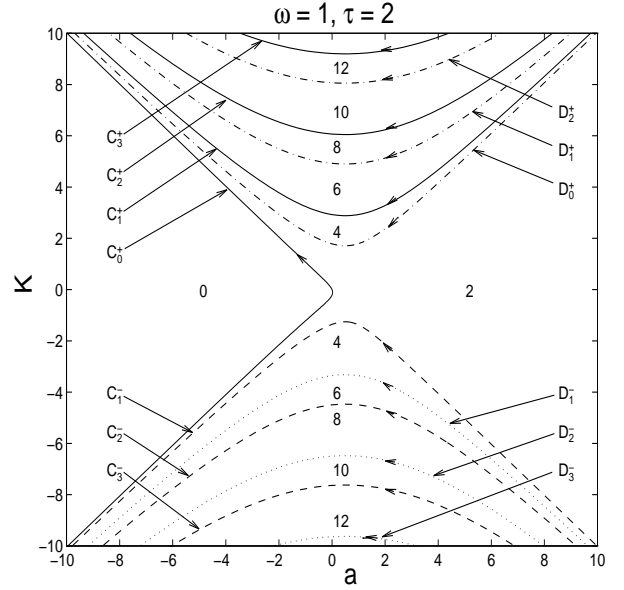,width=8cm,height=8cm}
\caption{%
Critical curves in (a,K) plane. The numbers indicate the number 
of eigenvalues in the right half plane of the complex 
eigenvalue space. The amplitude death region is the region where 
there are zero eigenvalues in the right half plane. The arrow
on each curve shows the direction along which $\beta$ increases.
}
\label{FIG:crit}
\end{figure}
and hence $\det(M) = \sin(\beta\tau)$. It is easily seen that
$\det M > 0$ on $C_n^+$, $D_n^+$, and $\det M < 0 $ on $C_n^-$,
$D_n^-$.  And finally we know that the stability region for
$\tau = 0$ is given by $a < K$. Combining these facts we present
our results for the
stability region and the number of eigenvalues in
Fig.~\ref{FIG:crit}. The {\it amplitude
death} region is the region where there are no (i.e. zero)
eigenvalues in the right half plane.
\subsection{Periodic Solutions}
\label{sec:per}
We now examine the region where the origin is unstable. For $\tau=0$
this region sustains periodic solutions as discussed in the 
introductory remarks of this section. We now look for periodic 
solutions in the presence of time delay. For this, it is convenient 
to cast Eq. (\ref{EQN:E3}) in polar form. For simplicity, we also
set $a=1$ implying that the oscillator without any kind of 
feedback has a unit circle as its periodic solution, and the phase 
increases linearly on the circle.  Writing Eq. (\ref{EQN:E3}) 
in polar coordinates we have,
\be
\label{EQN:E1r}
\dot{r}(t)  =  [1-r^2(t)]r(t) - K r(t-\tau) 
                 ~\cos[\theta(t-\tau)-\theta(t)] ,
\ee
\be
\label{EQN:E1theta}
\dot{\theta}(t)  =  \omega - K \frac{r(t-\tau)}{r(t)} 
~\sin[\theta(t-\tau)-\theta(t)] .
\ee
Recall that when $\tau = 0$, Eq. (\ref{EQN:E3}) has the 
time asymptotic periodic solution 
$Z(t) = \sqrt{1 - K} ~e^{i \omega t}$.
For non zero $\tau$ we can still assume a periodic solution of 
the form $Z(t) = R ~e^{i \Omega t}$; i.e.,
we are looking for solutions
of the form $r(t) = R$ and $\theta(t) = \Omega t$, where $R$ and 
$\Omega$ are real constants. In fact, it can be checked that this 
is the only solution which has a linear growth of the phase. 
Substituting this form in Eqs. (\ref{EQN:E1r}) and 
(\ref{EQN:E1theta}) and after some algebra, we obtain the following 
relations for the amplitude and the frequency of the oscillator:
\bea
\label{EQN:psamp}
&R & = \sqrt{ 1 - K ~\cos(\Omega \tau) } , \\
\label{EQN:psfre}
&\Omega & = \omega + K ~\sin(\Omega \tau) .
\eea
The oscillator can now lie
outside the unit circle when $\cos(\Omega \tau) < 0$; that is,
the amplitude of the limit cycle can increase beyond unity if
$\tau \in ( (2n+\frac{1}{2})\pi/\Omega,
(2n+\frac{3}{2})\pi/\Omega ) $.
However, the amplitude is bounded for
any value of $\tau$ 
because $max(R) \le \sqrt{1+K}$. 
So for any given $K$, the amplitude of the limit
cycle stays in the interval $[\sqrt{1-K},\sqrt{1+K}]$ for
arbitrary values of $\tau$. The above condition on $\Omega\tau$
can be used in Eq. (\ref{EQN:psfre}) to infer bounds on the
frequency of the limit cycle: $\omega - K \le \Omega \le
\omega+K$.

Eq. (\ref{EQN:psfre}) admits multiple solutions for the frequency 
$\Omega$.  The left-hand side, $y_1 = \Omega$, of (\ref{EQN:psfre}) 
is a straight line and the right-hand side, 
$y_2 = \omega + K ~\sin(\Omega\tau)$, is a sinusoidal curve 
with amplitude $K$ and a shift of $\omega$ above the
horizontal axis, $y = 0$. The multiple solutions (frequencies) are 
given by the intersection of the curves $y_1$ and $y_2$. As the 
value of $K$ is increased for a fixed value of $\omega$, 
the curve $y_2$ makes more and more intersections with $y_1$ and 
thus a set of {\it multiple frequencies} comes into existence.
Similarly a variation in $\omega$ will also bring about changes
in the number of possible solutions. But for all values of
$\omega$, including for $\omega = 0$,  multiple frequency solutions
are possible. The existence of multiple frequencies is a
characteristic feature of time delay systems and has been noted
before in the context of the Kuramoto model with time 
delay \cite{NSK:91} and in other studies \cite{SW:89,RSJ:98,YS:99}. 
These multiple frequency states that coexist with the lowest 
frequency state can be accessed by a suitable choice of initial 
conditions and
have potential applications in coupled oscillator models of the
human brain.  
\EM
\begin{figure}
\centerline{\psfig{file=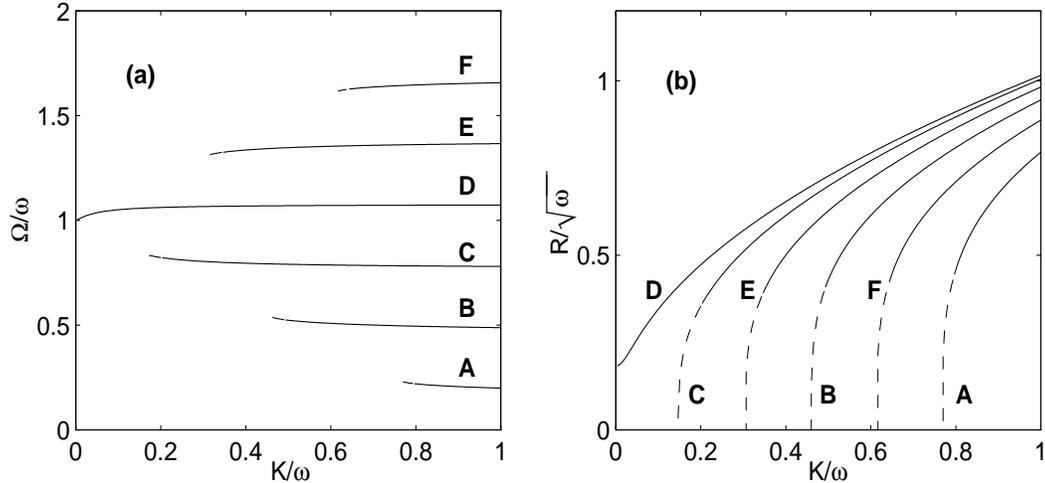,height=7cm,width=14cm}}
\caption{%
Finite time delay leads to the existence of multiple periodic
states, which can coexist with the primary limit cycle state. 
The frequencies and amplitudes of these states (for $\tau= 0.68$) are
plotted as a function of the coupling strength $K$ in
(a) and (b) respectively. As the value of the strength of 
the feedback, $K$, is increased, the number of periodic orbits 
increases.  In both the plots the dashed portions of the curves 
are unstable regions.
}
\label{FIG:multi}
\end{figure}
\begin{multicols}{2}
In Fig. \ref{FIG:multi}(a) we  plot these multiple frequencies,
$\Omega$, for $\omega = 30$ and 
$f = 3.25$ where $f = \omega\tau/(2\pi)$. 
Fig. \ref{FIG:multi}(b) shows the
corresponding amplitudes of the multiple states.  
Some of these states merge if $f$ is an integer.
This can be inferred from the intersections of
the amplitude curves in Fig. \ref{FIG:multi}(b).  At these points, the
oscillator has two frequencies with a single amplitude.  To find
out these frequencies, let $\Omega_1$ and $\Omega_2$ be the two
frequencies at these degenerate points. Substituting these
values in the expression for $R$ in (\ref{EQN:psamp}), we get
$\frac{\Omega_1}{\omega} = \frac{\Omega_2}{\omega} +
\frac{m}{f}$, where $m$ is an integer.

\begin{figure}
\PSFIG{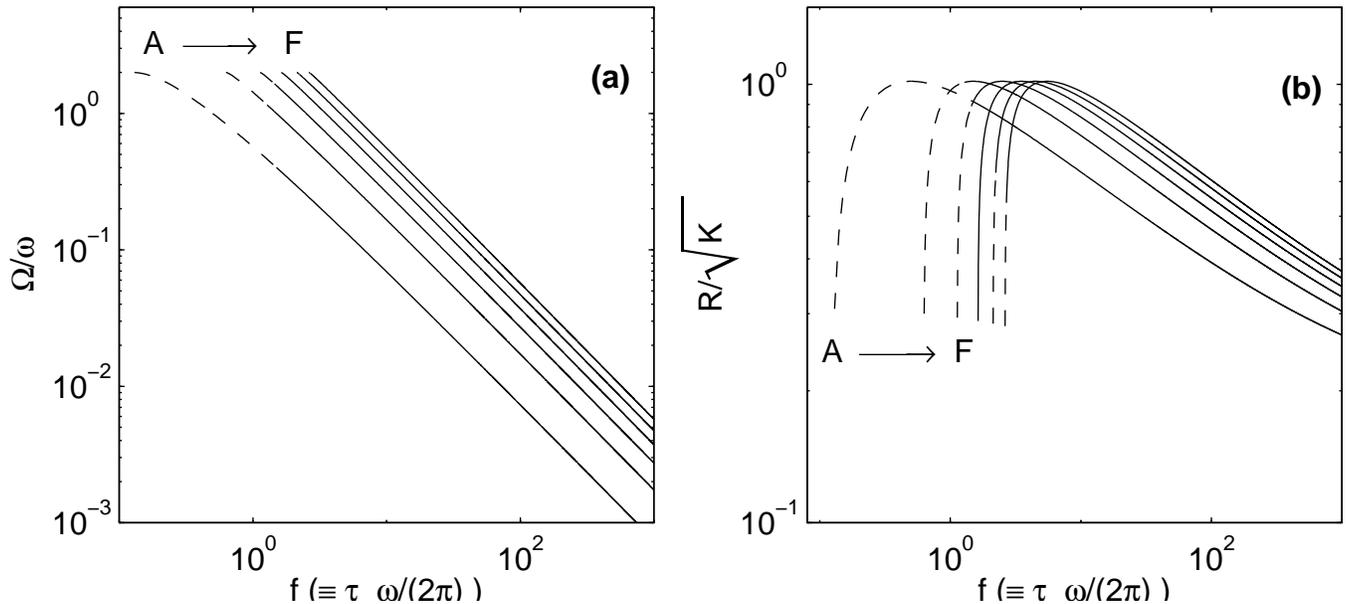}
\caption{%
The frequencies and amplitudes of the multiple periodic states
are plotted as a function of $f=\tau \omega /(2 \pi)$ in (a) and
(b) respectively at $K=\omega=30$.  At large values of $\tau$,
the frequency suppression is algebraic and is proportional to
$1/\tau$. In both the plots the unstable portions are indicated
by dashed lines.
}
\label{FIG:supp}
\end{figure}

{\it Stability of the Multiple States.}
The stability of these multiple periodic solutions can be obtained by 
linearizing about each of the solutions. The linearized matrix of
Eqs. (\ref{EQN:E1r}) and (\ref{EQN:E1theta}) about the periodic solutions
can be written as 
$$
M_1 ~= ~\left[\matrix{ 
A - B e^{-\lambda\tau}  &R C  (1- e^{-\lambda\tau}) \cr
-\frac{C}{R}(1-e^{-\lambda\tau}) &B (1-e^{-\lambda\tau}) } \right] ,
$$
where $A = 1 - 3 R^2,$ $B = K ~\cos(\Omega\tau),$ and 
$C = K ~\sin(\Omega\tau)$.
The corresponding eigenvalue equation is 
\hbox{$ \det(M_1 - \lambda I ) = 0$}, or 
$$ A_1 e^{-2\lambda\tau} - A_2 e^{-\lambda \tau} + A_3 +
\lambda^2 - A_4 \lambda + A_5 \lambda e^{-\lambda \tau} = 0,
$$
where $A_1 = B^2 + C^2$, $A_2 = AB+B^2-2C^2$, $A_3= AB + C^2$,
$A_4=A+B$ and $A_5=2B$.
If $\tau=0$, then $\lambda = \{0, -2(1-K)\}$.  
%
We thus recover the result that the periodic solution is stable
for $\tau = 0$ when $K < 1$. For $\tau \neq 0$ we need to solve
the eigenvalue equation numerically.
Our numerical results for the stability of the multiple periodic 
states are incorporated in Fig.~\ref{FIG:multi} and Fig.~\ref{FIG:supp}
where the dashed 
portions of the curves indicate unstable regions.
At large value of  $\tau$, the frequency of the oscillation , $\Omega$,
gets reduced. This is true for all the multiple frequency states
that the system possesses. 
In Fig.~\ref{FIG:supp}(a) the normalized frequency
of each of the states is plotted against the time delay on a
log scale. The frequencies are suppressed at a rate proportional
to $\tau^{-1}$. 
Fig.~\ref{FIG:supp}(b) shows the
corresponding amplitudes plotted against time delay. This feature of
{\it frequency suppression} has been observed in the past for large
coupled systems \cite{NSK:91} and once again seems to have its roots 
in the behavior of a single oscillator in the presence of a time 
delayed feedback drive. However, for short time delays, as may be
seen from the figures, the effect of time delay is some what distinct - 
both the frequency and the amplitude of the oscillator increase slightly
with $\tau$. To understand this behavior we present in 
Section~\ref{SEC:smalltau}
analytic expressions for the time evolution of $Z(t)$ in the limit of 
small time delay.

\subsection{Small $\tau$ approximation} 
\label{SEC:smalltau}
For short time delays, it is still worthwhile to 
expand the delay variable $Z(t-\tau)$ in a Taylor series
despite occasional warnings \cite{Dri:77book}.
Let us write
$$
Z(t-\tau) = Z(t) - \tau \dot{Z}(t) + \frac{\tau^2}{2} \ddot{Z}(t) \ldots 
$$
Substituting in Eq.~(\ref{EQN:E3}), 
the following two equations for the first two orders can be written. 
\bea
\label{EQN:E3one}
&{ }& O(\tau):~~ \nonumber \\
&{ }& \dot{Z}(t) = 
                \left[ (1 + i \omega - \mid Z(t) \mid^2) Z(t) 
                        - K Z(t) \right]\Big/(1-K\tau) , \\
\label{EQN:E3two}
&{ }& O(\tau^2):~~ \nonumber \\
&{ }& \ddot{Z}(t) 
  + b \dot{Z}(t) - c (1 - K + i \omega - \mid Z(t) \mid^2) Z(t) = 0 ,  
\eea
where $b = 2(1-K\tau)/K\tau^2$ and $c = 2/K\tau^2$.
It can be checked easily that the asymptotic periodic solutions 
are given by
\bea
O(\tau):~~ &{ }& Z(t) = \sqrt{1-K}~e^{i\omega^\prime t}, \\
O(\tau^2):~~ &{ }& Z(t) = \left[1-K+\frac{\omega^2 K 
             \tau^2}{2 (1-K\tau)^2}\right]^{1/2} ~e^{i\omega^\prime t},
\eea
where
$\omega^\prime = \omega / (1-K\tau)$. 
The effect of small $\tau$ is evident on 
the frequency and the amplitude: both show a slight rise.
\EM
\BM

\section{Nonlinear Feedback Effects}
\label{sec:nlin}

In this section we include the nonlinear feedback term ($K_2 \neq 0$)
and examine the dynamics of the limit cycle oscillator in the 
presence of the complete feedback term $f$ as given in (2). 
We then have
\bea
\label{EQN:E1non}
\dot{Z}(t) = (a + i \omega - \mid Z(t) \mid^2) Z(t) 
             &-& K_1 Z(t-\tau) \nonumber \\
             &-& K_2 Z^2(t-\tau) .
\eea
The transformations $Z(t)\rightarrow e^{i\pi} Z(t)$ 
and $K_2 \rightarrow -K_2$ leave the
equation unchanged. The bifurcation diagrams, 
thus, appear symmetric about 
$K_2 = 0$ axis with the exception that the orbits are rotated by 
an angle of $\pi$ about the origin.

\subsection{$\tau = 0$}
In order to distinguish the additional effects arising from the 
nonlinear feedback term we first turn-off the time delay and examine 
the dynamics for $\tau =0$. These solutions also correspond to the 
special class of solutions with time delay when the solutions have 
a  periodic recurrence with a period $\tau$. 
These are generally termed as {\em phase trapped} solutions in 
the literature. 
Let $Z = X+iY$. The evolution equations are
\bea
\label{EQN:xtau0}
\dot{X} &=& (a-X^2-Y^2)X-\omega Y - K_1 X - K_2 (X^2-Y^2), \\
\label{EQN:ytau0}
\dot{Y} &=& \omega X + (a-X^2-Y^2)Y - K_1 Y - 2 K_2  X Y .
\eea

\subsubsection{Fixed points}
The fixed points of the system are found by solving
$\dot{X}=0$ and $\dot{Y}=0$ simultaneously. 
In contrast to the earlier case of a linear feedback drive, 
there are now three 
fixed points of the system including the origin.
In Table 1, the fixed points and expressions for
the eigenvalues of linear perturbations around these fixed
points are listed using the definitions
\begin{minipage}{\columnwidth}
\begin{table}[!h] 
\caption{}
\begin{tabular}{cccc}
\hline
Symbol & index &   Fixed point & Eigenvalues \\
                      & j  & $(X_j^*,Y_j^*)$ & $\lambda$ \\
\hline
$D$    &   0 & $(0,0)$    & $-\tilde{K}_1 ~\pm~ i \omega$ \\
$FP_1$ &   1 & $(-\frac{K_2}{2}+\sqrt{G},\frac{\omega}{K_2})$ & 
$\tilde{K_1} ~\pm~T_1$ \\
$FP_2$ &   2 & $(-\frac{K_2}{2}-\sqrt{G},\frac{\omega}{K_2})$ & 
$\tilde{K_1} ~\pm~T_2$ \\
\hline
\end{tabular}
\end{table}
\end{minipage}

\begin{figure}
\PSFIG{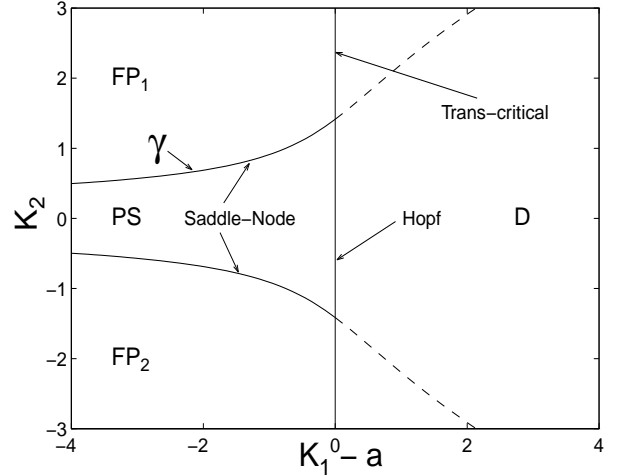,height=6.5cm,width=8cm}
\caption{%
The bifurcation diagram of Eq. (\protect\ref{EQN:E1non}) for $\tau=0$
in $(\tilde{K}_1,K_2)$ plane for $a=1$, $\omega = 1$. $D$ is the
region of amplitude death, $FP_1$ and $FP_2$ are the regions of
stability of the fixed points $FP_1$ and $FP_2$ respectively.
The periodic solution is stable in the region $PS$.
}
\label{FIG:bdnodelay}
\end{figure}

\begin{figure}
\PSFIG{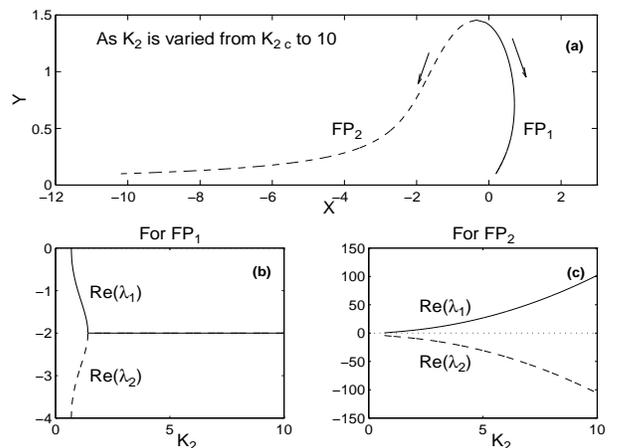,height=6.0cm,width=8cm}
\nopagebreak
\caption{%
The movement of the fixed points of Eq. (\protect\ref{EQN:E1non})
for $\tau = 0$ ($K_1 = -1$, $a=1$, and $\omega=1$) is shown in (a)
as $K_2$ is varied from the critical value $K_{2 c} = 0.6871$
to $10$. The behavior of the real parts of
the eigenvalues for the corresponding fixed points are plotted in
(b) and (c) respectively and show that $FP_1$ is stable and
$FP_2$ is unstable in this case.
}
\label{FIG:ev}
\end{figure}

$G = (\frac{K_2}{2})^2 - \tilde{K_1} - \frac{\omega^2}{K_2^2}$,
$\tilde{K}_{1} = K_{1} - a$
and $T_i = \sqrt{({X_i^*}^2+{Y_i^*}^2)^2-\omega^2}$.
As can be seen, the origin exists as a fixed point for all the
values of the parameters $K_1, K_2, a $ and $\omega$. However, it is
stable only when $\tilde{K}_1 > 0$, i.e. when $K_1 > a$.  The
other two fixed points $FP_1$ and $FP_2$ exist in the region where 
$G > 0$. This condition gives rise to $K_2  > \gamma(\tilde{K}_1)$
and $K_2 < - \gamma(\tilde{K}_1)$, where  $\gamma(\tilde{K}_1) 
\equiv \sqrt{2} \sqrt{\tilde{K}_1 + \sqrt{\tilde{K}_1^2 + \omega^2}}$.
This region overlaps with $\tilde{K}_1 > 0$.
Fig.~\ref{FIG:bdnodelay} shows the bifurcation diagram. 
The regions marked with
$D$, $FP_1$ and $FP_2$ are the regions in which the corresponding 
fixed points are stable. The region $D$ is bounded on the right hand 
side by $K_1 = a$. The region $FP_1$ is bounded by the curves 
$\tilde{K}_1 = 0$
and $K_2 = \gamma(\tilde{K}_1)$. The region $FP_2$ is bounded by
the curves $\tilde{K}_1 = 0$ and $K_2 = -\gamma(\tilde{K}_1)$.
The movement of these fixed points and the corresponding eigenvalues 
are shown in Fig.~\ref{FIG:ev}. 
\newpage
\begin{figure}
\PSFIG{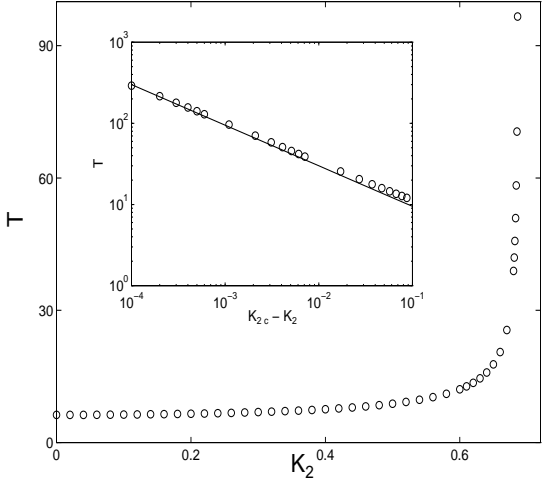,height=7cm,width=8cm}
\caption{%
The variation of the period of the limit cycle in the region
$PS$ of Fig.~\ref{FIG:bdnodelay}.
as a function of $K_2$ for a fixed value of $K_1=-1$.
As $K_2$ is increased from $0$ to $\gamma(K_1)$ the period
increases from $2\pi/\omega$ to infinity with a scaling proportional
to $1/\protect\sqrt{(K_{2 c} - K_2)}$
and leading to a saddle-node
bifurcation. The inset displays on a log scale the behavior close to
the critical region. The solid curve
is a plot of $3/\protect\sqrt{K_{2c}-K_2}$.
}
\label{FIG:tp}
\end{figure}
\subsubsection{Periodic orbit}
Eqs.~(\ref{EQN:xtau0}) and (\ref{EQN:ytau0}) have a 
stable periodic solution in the region 
$-\gamma(K_1) < K_2 < \gamma(K_1)$. 
The periodic orbit develops a saddle $(FP_2)$ and a node
$(FP_1)$ as $K_2$ is increased above $\gamma(K_1)$. Similarly as 
$K_2$ is decreased below $-\gamma(K_1)$ a node $(FP_2)$ and a 
saddle $(FP_1)$ are born.
The periodic orbit is a circle for $K_2 = 0$ and the oscillator 
moves linearly on the circle with a frequency $\omega$. 
As $K_2$ is increased or decreased the period of the limit cycle 
increases and tends to infinity on the curve 
$K_2 = \pm \gamma(\tilde{K}_1)$
giving birth to a saddle-node point. 
In Fig.~\ref{FIG:tp} we have plotted the period of the limit cycle 
as a function of $K_2$ for $K_1 = -1, a = 1, \omega = 1$.
The inset of this figure also shows the scaling of the approach
to infinite time period which in this case is the typical inverse 
square root scaling of the saddle-node bifurcation phenomenon.
As can be seen from Table 1, one of the eigenvalues of the Jacobian
of Eqs.~(\ref{EQN:xtau0}) and (\ref{EQN:ytau0}) 
is zero at the critical value
of $K_2$. The existence of the saddle-node bifurcation can thus be
rigorously established by reducing the dynamics 
to the center manifold \cite{GH:83}.

{\it Phase slips}.
As the magnitude of $K_2$ is increased from $0$ to $\pm\gamma(K_1)$ 
the evolution of the phase of the oscillator takes on 
a nonlinear character. 
The parabolic nature of $\dot\theta(\theta)$ at its minimum
and its sinusoidal nature for other values of $\theta$ 
indicate the existence of a nonuniform motion of the oscillator
from $0$ to $2\pi$. 
The motion becomes jerky as the parameter is increased to bring a 
tangency at which point $\dot\theta = 0$. 
The time period becomes singular at the point of tangency. 
\begin{figure}
\PSFIG{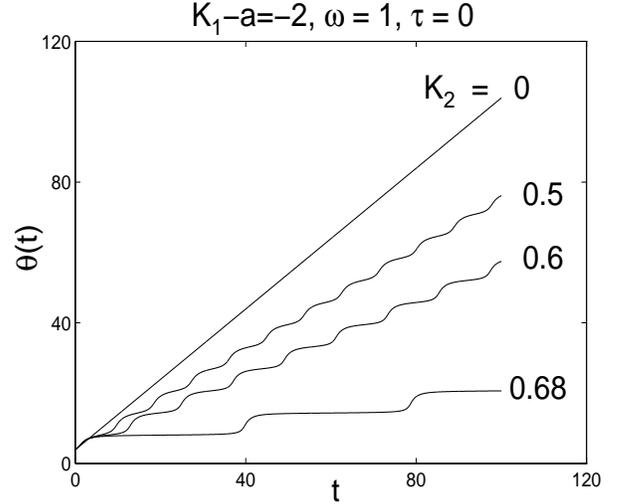,height=7cm,width=8cm}
\caption{%
Phase slips of the limit cycle oscillator for
$\tau = 0, K_1 = -1, a = 1, \omega = 1$.
The phase of the limit cycle oscillator evolves linearly on the circle
when there is no nonlinear feedback. With increasing $K_2$ one clearly
sees the phase slips suffered by the oscillator.
}
\label{FIG:pslip}
\end{figure}
It begins to spend more and more time in the vicinity of 
$\theta = \theta^*$ and less and less time at other values.  
Close to the curves $\pm \gamma(K_1)$ the oscillator 
shows clear phase slips of $2\pi$ as seen in Fig.~\ref{FIG:pslip}. 
This phase slip ends with phase quenching on
the boundary of $\pm \gamma(K_1)$.  In this state the phase of the
oscillator is a constant and such a state is often  
referred to as {\it phase death}. The phenomenon of phase slips 
plays an important role in the process of transition towards
a synchronized state in a system of a large number of coupled limit
cycle oscillators and has been the subject of some recent 
investigations \cite{ZHH:98}.

\subsection{$\tau > 0$}
We now return to our full feedback model and include 
the effect of finite time delay on the dynamical characteristics of 
the oscillator. Since we now have an infinite dimensional system,
the phase space dynamics can in general be quite complicated
and changes significantly as a function of the time delay parameter 
$\tau$.  
For our further analysis and description, we write down the 
Eq.~(\ref{EQN:E1non}) in polar coordinates.
\bea
\label{EQN:rnon}
\dot{r}(t) &=& [a-r^2(t)] r(t) 
- K_1 r(t-\tau) \cos[\theta(t-\tau)-\theta(t)] \nonumber \\
&{ }& \hspace{1cm} - K_2 r^2(t-\tau)~\cos[2\theta(t-\tau)-\theta(t)], \\
\label{EQN:thetanon}
\dot{\theta}(t) &=& \omega 
- K_1 \frac{r(t-\tau)}{r(t)} \sin[\theta(t-\tau)-\theta(t)] \nonumber \\
&{ }& \hspace{1cm} - K_2 \frac{r^2(t-\tau)}{r(t)} 
                      ~\sin[2\theta(t-\tau)-\theta(t)] .
\eea
\begin{figure}
\PSFIG{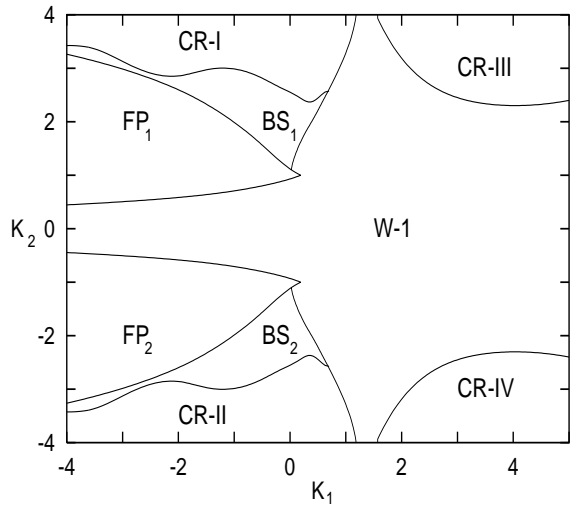,angle=270,height=7cm,width=8cm}
\caption{%
Bifurcation diagram of
Eq.~(\protect\ref{EQN:E1non}) for $\tau = 0.5$, $a = 1$, $\omega = 1$.
The areas marked
$FP_1$ and $FP_2$ refer to stability domains of these fixed points,
the region $W-1$ sustains limit cycle orbits encircling the origin
(winding number 1), the regions $BS_{1}, BS_{2}$ contain combinations of
{\it phase reversing}, {\it radially trapped} or {\it spiraling}
solutions in a birythmic existence,
and the regions marked $CR-I,CR-II,CR-III,CR-IV$ exhibit chaotic
behavior interspersed with periodic windows.
}
\label{FIG:bddelay}
\end{figure}
In the following we describe several different solutions 
which are found numerically (depicted in Fig.~\ref{FIG:bddelay}
and the following figures) 
and provide explanation in terms of the behavior of the 
amplitude and the phase. 
These solutions include birhythmicity, phase reversals, phase slips, 
radially trapped orbits, phase jumps, and spiraling orbits.  For all the
numerical integrations, unless stated otherwise, 
we use a constant history function: $Z(t) = Z(0), ~t\in [-\tau,0)$ 
and a constant step size (h) of integration. 
Several values of $h$ were tested for the accuracy of the results. 
Issues relating to the numerical 
integration of delay differential equations 
can be found in \cite{BPW:94}. 
In all our later results we set the parameters $a = 1$ 
and $\omega = 1$.

Before we begin our description of the solutions to 
Eqs.~(\ref{EQN:rnon}) and (\ref{EQN:thetanon}), 
we wish to point out an important 
connection between the present model and some of the other models 
that have been studied in the literature. 
Under the assumptions that $K_1 = 0$, 
\hbox{$\mid\theta(t-\tau)-\theta(t)\mid << 1$},
and \hbox{$r(t) = r(t-\tau) = 1$}, 
Eq.~(\ref{EQN:thetanon}) yields
\bea
\dot\theta(t) = \omega &-& K_2~\sin\theta(t-\tau) \nonumber \\
&-& K_2~[\theta(t-\tau)-\theta(t)]~\cos\theta(t-\tau).
\eea
The above equation with the first two terms on the right hand side 
describes a first order phase locked loop with 
time delay \cite{WWPOG:94}.
It also has the structure of the equation that can be used to model 
certain visually guided movements of biological limbs \cite{TKRGH:96}.

\subsubsection{Stability of the fixed points}
We proceed first to carry out a linear perturbation 
analysis of Eq.~(\ref{EQN:E1non})
around the three fixed points and then construct the 
bifurcation diagram for fixed values of $\tau$ through detailed 
numerical investigations.
The analysis of the stability of the origin remains unchanged by the
presence of the $K_2$ term since its contribution vanishes in the
linear limit.  So its behavior can be discerned from the previous
section (e.g. see Fig. \ref{FIG:death}(a)), 
where we saw that the amplitude death
region shrinks and moves to the right of the curve $\tilde{K}_1
= 0$ as $\tau$ is increased from $0$ and vanishes at a certain
critical value.  Depending on the strength of $\omega$, it may
reappear.  For the present study we choose $\omega = 1$
and $\tau = 0.5$, which is much less than the intrinsic time
period $2\pi$. For this value the amplitude death region
disappears and the origin is always unstable.  The fixed points
$FP_1$ and $FP_2$ are again the same as discussed in the
previous section, but their stability properties are now
significantly modified due to finite time delay effects.  Let
$Z^* = (X^*+iY^*)$ be one of the non-zero fixed points of the
system.  A linearization about $Z^*$ of (\ref{EQN:E1non}) yields
\bea
&{ }& \dot{z}(t) = (a+i\omega-2\mid Z^*\mid^2) z(t) \nonumber \\
&{ }&  \hspace{2cm} - {Z^*}^2 \bar{z}(t) - (K_1 +2 K_2 Z^*) z(t-\tau),
\eea
where $z=Z - Z^{*}$ and $\bar{z}$ is the complex conjugate of $z$. 
Write $z(t) = x(t) + i y(t)$
and assume each component to vary as $e^{\lambda t}$, where $\lambda$
is the eigenvalue of the linearized matrix, $M_2$,  
whose
eigenvalue equation is given by 
$\det(M_2 - \lambda I) = 0$, and can be written as
\bea
(A-B e^{-\lambda\tau} - \lambda)^2 
- 2({X^*}^2+{Y^*}^2) (A-B e^{-\lambda\tau}-\lambda) \nonumber \\
+ (2 K_2 Y^* e^{-\lambda\tau} - \omega)^2 = 0 ,
\label{EQN:E1nonev}
\eea
where  
$A = a - ({X^*}^2 + {Y^*}^2)$ and $B = K_1 + 2K_2 X^*$.
The evolution of $\lambda$ in the complex plane as $K_1$ and $K_2$ 
are varied decides the stability of the fixed points. The relation 
between $K_1$ and $K_2$ for the critical boundary is not very 
transparent from the above equation. We use Eq. (\ref{EQN:E1nonev}) to 
determine the boundaries of the regions of these 
fixed points numerically
and present the bifurcation diagram in Fig.~\ref{FIG:bddelay}.
The regions labelled $FP_1$ and $FP_2$ represent the regions of 
stability of the respective fixed points.
When $K_2$ is small and one is in the region where the origin is 
unstable but $FP_1$ and $FP_2$ are stable, the limit cycle orbit 
preserves its identity and the results are similar to those discussed
in earlier sections. As $K_2$ is increased in magnitude, however,
several interesting new orbits depicted in the bifurcation diagram 
appear. We discuss these features below.

\subsubsection{Periodic solution}
\begin{figure}
\PSFIG{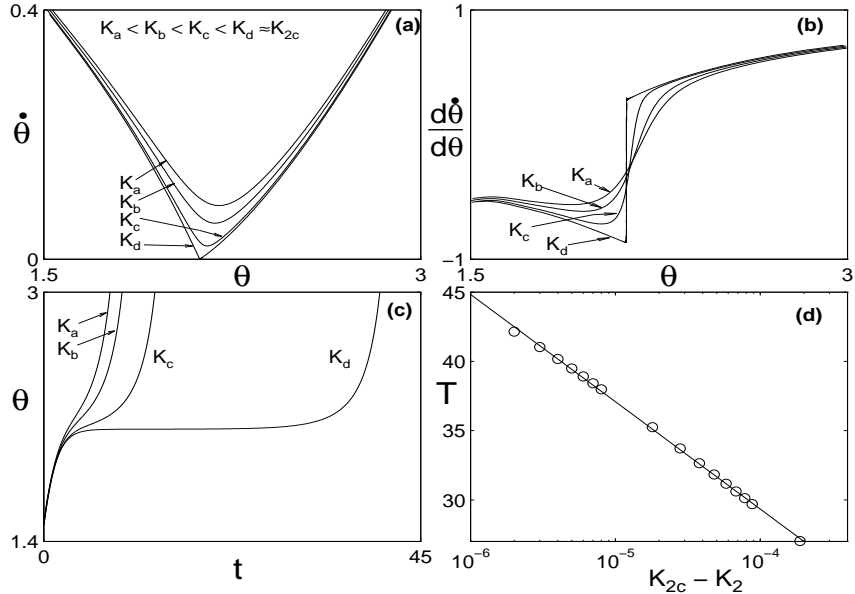,height=8cm,width=12cm}
\caption{%
The nature of (a) $\dot{\theta}$ and (b) $d\dot\theta/d\theta$ 
as a function of $\theta$ close to the bifurcation point, and 
(c) the corresponding stretching of the time period of the periodic 
orbit. (d) A semilog plot of the numerically obtained time periods
$T$ vs. $(K_{2c}-K_2)$ (shown as circles) which indicates a 
logarithmic scaling of the singularity, {\em viz.} 
$T = -3.3665 \log(K_{2c}-K_2) - 1.6772$ (solid line).
Here $K_{2c} = 0.923488$, $K_1 = 0$, $a = 1$, $\omega = 1$ and
$\tau = 0.3$.
}
\label{FIG:disc}
\end{figure}

The simple periodic orbit (limit cycle) exists for all small values 
of $K_2$ in the region marked as W-1 in Fig.~\ref{FIG:bddelay}. 
As $K_2$ is increased this stability can be lost through a variety 
of bifurcations (e.g. saddle node bifurcation, bistability or chaos). 
The most interesting effect of a finite delay time appears to be on 
the nature of the saddle-node bifurcation.
Although the period of the limit cycle becomes infinite at the 
bifurcation point, the scaling behavior and the nature of the 
orbit dynamics near this point are 
significantly different from those of the $\tau = 0$ case. 
To illustrate this point, we have plotted in Fig.~\ref{FIG:disc}(a)
a set of $\dot\theta$ vs. $\theta$ curves for $\tau = 0.3, K_1 = 0$ 
as $K_2$ is successively increased towards
the critical value. Note that the nature of the $\dot\theta$ curve 
is not parabolic near the critical point and has distinct asymmetries. 
Further, $d\dot\theta/d\theta$ develops a discontinuity 
at this point as shown in Fig.~\ref{FIG:disc}(b).
Since $\dot\theta$ goes to $0$ near the critical point, the period 
of the limit cycle keeps increasing as $K_2$ approaches the critical 
value as shown in Fig.~\ref{FIG:disc}(c). The discontinuity of 
$d\dot\theta/d\theta$ is a significant deviation from the standard 
conditions of a saddle-node bifurcation \cite{GH:83} and brings about 
the change in the scaling behavior.  To analytically estimate the 
time period close to the bifurcating point,
the $\dot\theta$ curve, in the vicinity of the critical point, 
can be approximated as
\EM
\begin{figure}
\PSFIG{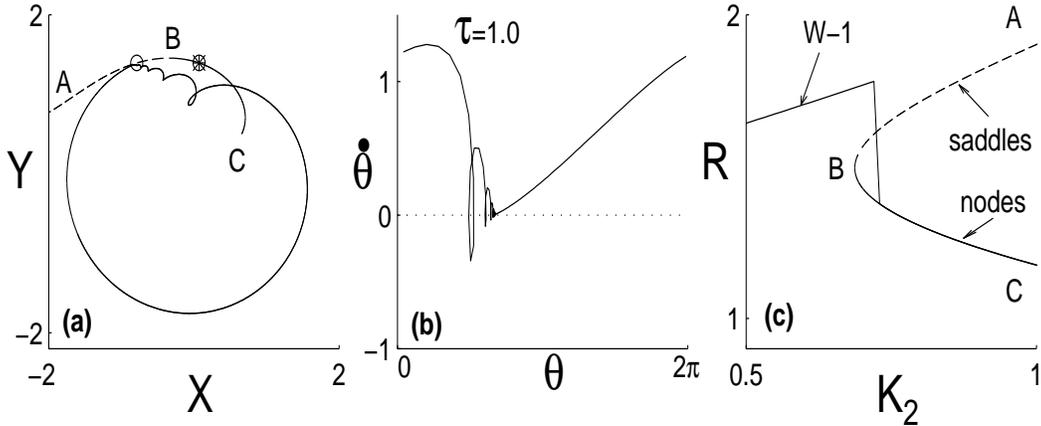,height=6cm,width=14cm}
\caption{%
Periodic orbit losing stability by colliding with the saddle.
(a) periodic orbit just before the critical value of $K_2$,
(b) corresponding evoluation of $\dot\theta$ as a function of $\theta$.
(c) shows the bifurcation diagram as $K_2$ is increased. $R$ is the
maximum value of the asymptotic state of the oscillator.
The parameters are $\tau = 1.0, K_1 = -1, a =1$, and $\omega = 1$.
}
\label{FIG:collision}
\end{figure}
\BM
\be
\dot{x} = \cases{ -b_1 x + c_1, & if $ x  < 0 $ \cr
                     b_2 x + c_1, & if $ x > 0 $}
\ee
where $x = \theta - \theta^*$, $\dot\theta$ is minimum at $\theta^*$
for a given value of $K_2$, and $b_1, b_2$ and $c_1$ are all positive 
constants.  ($c_1 = 0$ when $K_2 = K_{2c}$).
The time period across a thin region $2\delta$ around $x = 0$ is now 
given by 
\bea
T & \approx & \int_{-\delta}^{0} \frac{dx}{-b_1 x + c_1} + 
            \int_{0}^{\delta}{\frac{dx}{b_2 x + c_1}}  \nonumber \\
& = &  - \Big(\frac{1}{b_1} +  \frac{1}{b_2} \Big) \log c_1 \nonumber \\
&{ }& \hspace{1cm} + \log\Big( (b_1\delta+c_1)^{1/b_1} 
                                 (b_2\delta+c_1)^{1/b_2} \Big)
\eea
As seen from the first term in the expression above, 
the time period goes to infinity as $c_1\to 0$ with a 
logarithmic scale. This is in contrast to the 
inverse square root scaling for the $\tau= 0$ case. 
The above scaling agrees quite well with our numerical results
as shown in Fig.~\ref{FIG:disc}(d).
The actual mechanism of the loss of stability of the periodic orbit 
is through a collision with the saddle point. 
This is illustrated in Fig.~\ref{FIG:collision} which is plotted for
$K_1 = -1$ and $\tau = 1.0$.
The curves BC and BA are the stable ($FP_1$) and unstable ($FP_2$) 
branches,
and they are the same as those depicted in Fig.~\ref{FIG:ev}(a).
As seen in the $(X,Y)$ plane in Fig.~\ref{FIG:collision}(a),
the orbit collides with an existing saddle point and is kicked to the
stable node. Fig.~\ref{FIG:collision}(b) shows the corresponding 
behavior of $\dot\theta$ vs. $\theta$ just before the critical value 
of $K_2$.  Notice that its intersection with the $\dot\theta = 0$ line 
is non-parabolic.  There are also additional intersections which 
however are not fixed points
since $\dot{r} \ne 0$ at these points. These loops have significance for
the phase reversal orbits which we discuss in Section~\ref{SEC:pr}.
In Fig.~\ref{FIG:collision}(c), $R$ is the maximum amplitude of the 
asymptotic state of the orbit at each $K_2$.

\subsubsection{Phase reversals}
\label{SEC:pr}
\begin{figure}
\PSFIG{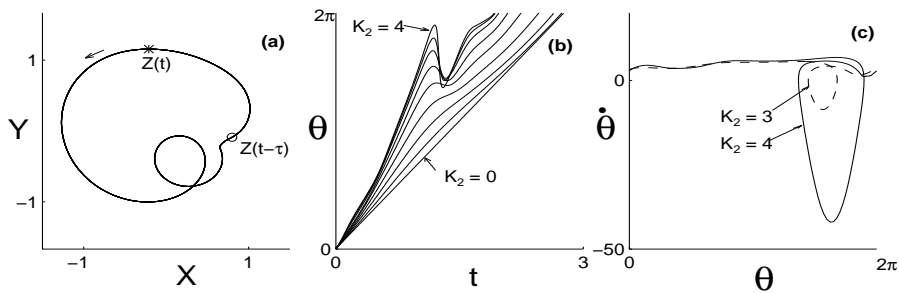,width=12cm,height=4cm}
\caption{%
Phase reversal.
(a) A phase reversing orbit of the limit cycle in the $X-Y$ space for 
$K_2 = 4.0$.
(b) Temporal evolution of the phase as measured from the origin as a 
function of
$K_2$ as $K_2$ is increased from $0$ to $4$ in steps of $0.5$.
(c) The multivaluedness of $\dot\theta(\theta)$.
The other constants are $K_1 = 1.4, a = 1, \omega = 1$, 
and $\tau = 0.5$.
}
\label{FIG:preversal}
\end{figure}

While still inside the region W-1,
the periodic orbit can show a reversal of its phase with time with
the same period as that of the orbit;
the curve $\dot{\theta}(\theta)$ develops a fold as 
$K_2$ is increased. Depending on the value of $K_1$ there can be 
one or more than one folds. 
The acquired additional loop as shown in Fig.~\ref{FIG:preversal}(a) is
not around the origin; the winding number of the 
periodic orbit continues to be one. However the phase of the orbit as
measured from the origin undergoes a reversal in that region
(Fig.~\ref{FIG:preversal}(b)), in contrast with the phase slip
behavior discussed earlier. 
This is purely the effect of the time delay appearing in the nonlinear 
feedback term.  The phase reversing orbits are prominent at the 
boundaries of the W-1 region.
Such periodic orbits for a single
oscillator have been observed for externally driven systems and
basically arise due to the excitation of higher harmonics from
the resonant interaction of the external driver with the basic
oscillator \cite{SSS:83}. 
They cannot exist for a single autonomous 
oscillator in the absence of time delay due to dimensional constraints.
However the presence of time delay in our autonomous model
increases the dimensionality of the system and hence permits the
existence of such orbits. 
This is a new kind of orbit which does
not seem to have been noticed or discussed before in the context
of large systems of coupled oscillators and it would be
worthwhile looking for their existence in such systems. \\

\subsubsection{Effect of time delay on phase slips}
\begin{figure}
\PSFIG{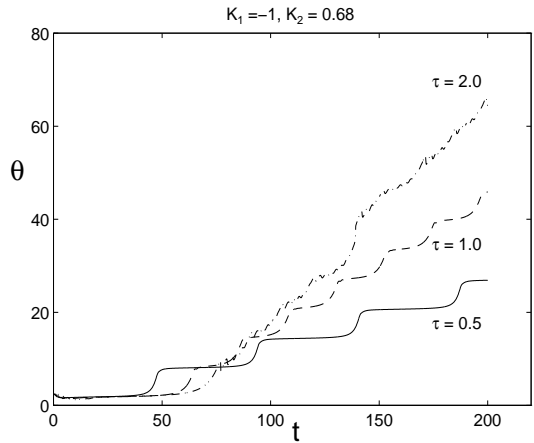,height=6cm,width=7cm}
\caption{%
The effect of time delay on phase slips.
($K_2 = 0.68$, $K_1 = -1, a = 1$, and $\omega = 1$.)
}
\label{FIG:pslipd}
\end{figure}

The phenomenon of phase slips continues to exist for non zero
values of time delay as well. To study the effect of time delay
on the phase slips, we examine Fig.~\ref{FIG:pslip} and choose the
parametric values corresponding to the phase slips shown by
the bottom most curve, i.e. for $K_2 = 0.68$, and introduce finite
time delay. The results are shown in Fig.~\ref{FIG:pslipd}. Time
delay has the effect of reducing the sharpness of the phase slips
and, at the same time, increasing the angular speed as seen by the
slope of $\theta$. For longer time delays, (e.g. for $\tau = 2.0$
in the figure), phase slips are less evident and in fact the orbits
exhibit chaos.
\EM
\begin{figure}
\PSFIG{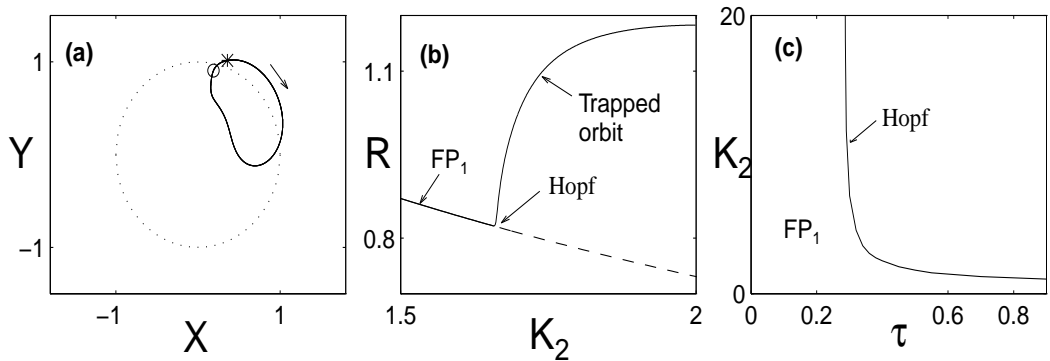,height=5cm,width=14cm}
\caption{%
Radial trapping.
(a) A radially trapped orbit of the limit cycle in the $X-Y$ space.
$\star$ and $\circ$ show two points separated in time by $\tau$.
(b) The bifurcation diagram plotted as a function of $K_2$.
$R$ is the maximum amplitude of the asymptotic state of the oscillator.
(c) The bifurcation diagram in $(\tau,K_2)$ space.
The fixed point $FP_1$ loses stability in a Hopf bifurcation as the 
curve is crossed to the right side.
The other parameters are $\tau = 0.5$, $K_1 = -0.6, a = 1$, 
and $\omega = 1$.
}
\label{FIG:rtrapped}
\end{figure}
\begin{figure}
\PSFIG{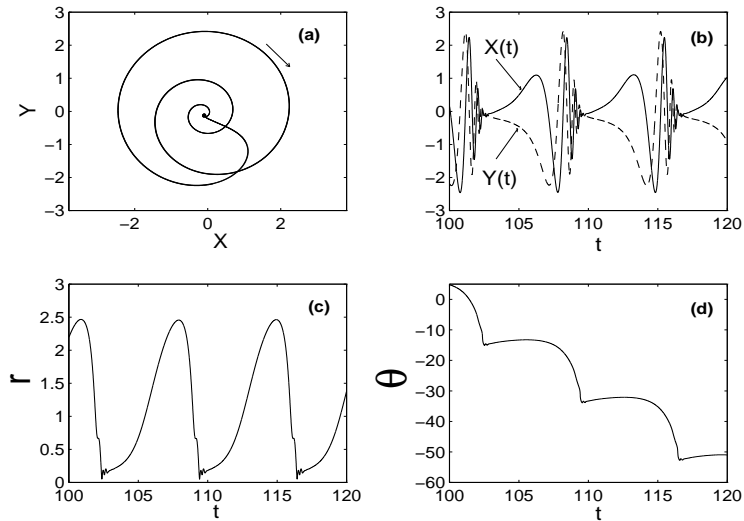,height=7cm,width=10cm}
\caption{%
A spiraling orbit.
The orbit of the limit cycle in the $X-Y$ space (a),
the temporal evolution of the individual components $X$ and $Y$ (b),
the amplitude $r$ (c), and that of the phase (d)
as measured from the origin are plotted for $K_2 =
-3., K_1 = 0.2, a = 1, \omega = 1$, and $\tau = 0.5$.
}
\label{FIG:sp}
\end{figure}

\BM
\subsubsection{Radially Trapped Solutions}
As we cross the boundary of $FP_1$, 
the node $FP_1$ loses stability in a Hopf bifurcation
as $K_2$ is increased resulting in a periodic orbit encircling
$FP_1$. For negative $K_2$, the corresponding node is $FP_2$.
These orbits have winding number zero around the origin and exist
in the regions $BS_1$ and $BS_2$ bordering those marked $FP_1$ 
and $FP_2$.  We call them radially trapped solutions because, 
viewed from the origin, 
they are restricted to a region of phase space and seem to oscillate 
within a restricted physical space. 
One such radially trapped orbit is shown in Fig.~\ref{FIG:rtrapped}(a)
with the corresponding bifurcation 
diagram in Fig.~\ref{FIG:rtrapped}(b).
The oscillator moves clockwise. Increased delay requires only weaker 
feedback to destabilize the node but the value saturates as shown 
in Fig.~\ref{FIG:rtrapped}(c); the curve
shown is plotted using Eq.~(\ref{EQN:E1nonev}) and is 
verified numerically.

\subsubsection{Spiraling solutions and phase jumps}
Another interesting periodic orbit that we find is shown in
Fig.~\ref{FIG:sp}(a) and can best be 
described as a {\it spiraling orbit}  since the amplitude of the
limit cycle (in the $(X,Y)$ space) first spirals out and then
comes back to its original state. The phase changes have a step
like character and often exhibit large jumps. Such orbits exist in 
a thin region near the radially trapped solutions.

\subsubsection{Birhythmicity}
\begin{figure}
\PSFIG{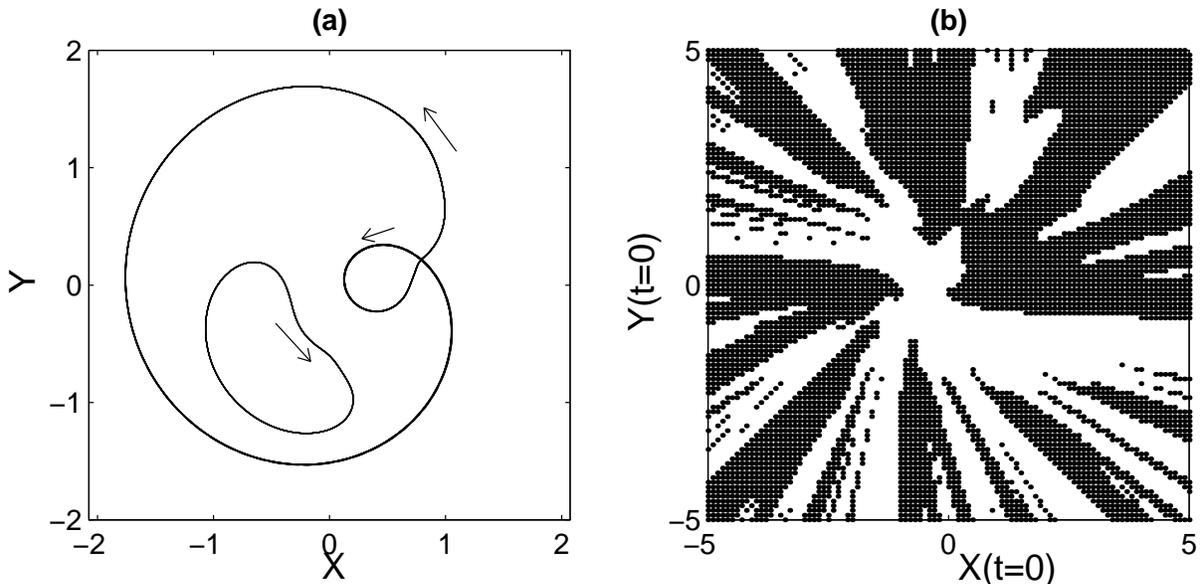,height=8cm,width=16cm}
\caption{%
Birhythmicity between two periodic orbits.
(a) Radially trapped and the phase reversal solution
coexisting for $K_2 = -1.6$, $K_1 = -0.2$, $a = 1$, $\omega = 1$, and
$\tau = 0.5$.
(b) A plot of the basins of attraction for these two states in the
space of the initial conditions $(X(0),Y(0))$.
The dotted region is the basin of attraction for the phase
reversing solution and the white region is that for the
radially trapped solution. To generate this figure the initial
conditions chosen were
$X(\theta) = X(t=0), \theta\in[-\tau,0)$,
and
$Y(\theta) = Y(t=0), \theta\in[-\tau,0)$.
}
\label{FIG:br}
\end{figure}

The {\it phase reversing} and the {\it radially trapped}
solutions can also coexist in the same parameter regime
(e.g. $K_2 = -1.6$, $K_1 = -0.2$, $a = 1$, $\omega = 1$, $\tau=0.5$ as 
shown
in Fig.~\ref{FIG:br}(a)) and thus exhibit a {\it birhythmic behavior}.
In fact, birhythmicity appears to occur also for the spiraling and
trapped solutions and is spread out over a large region of the
parameter space inside $BS_{1}$ and $BS_{2}$ in
Fig.~\ref{FIG:bddelay}. Birhythmicity is a common phenomenon in 
many biological cell
models, and is currently the subject of many studies \cite{Gol:96}.
The switching between the two states can take place by the slightest
perturbation to the initial states $(X(0),Y(0))$.
A plot of the basins of attraction for the two states of
Fig.~\ref{FIG:br}(a)  is shown in Fig.~\ref{FIG:br}(b) in the
space of the initial conditions $(X(0),Y(0))$.
The dotted region is the basin of attraction for the {\it phase
reversing} solution and the white region is that for the 
{\it radially trapped} solution. To generate this figure the initial 
conditions chosen were 
$X(\theta) = X(t=0), \theta\in[-\tau,0)$,
and 
$Y(\theta) = Y(t=0), \theta\in[-\tau,0)$. 
\EM
\BM
\hbox{ }
\begin{figure}
\PSFIG{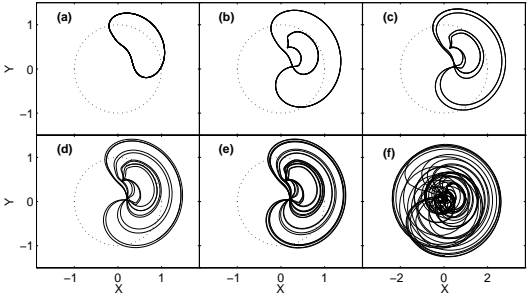,height=4cm,width=7cm}
\caption{%
The radially trapped orbit of the limit cycle can undergo a period
doubling sequence to the chaotic state. Some of the
periodic orbits with
(a) Period-1, $K_2 = 1.6$, (b) Period-2, $K_2 = 2.1$,
(c) Period-4, $K_2 = 2.13$, (d) Period-8, $K_2 = 2.18$,
(e) Period-16, $K_2 = 2.186$, and (f) a chaotic orbit, $K_2 = 2.6$
are shown. The other constants are
$K_1 = -0.2$, $a = 1$, $\omega = 1$, and $\tau = 0.5$.
The scale of (d) is twice that of the other plots.
}
\label{FIG:ptb}
\end{figure}
\begin{figure}
\PSFIG{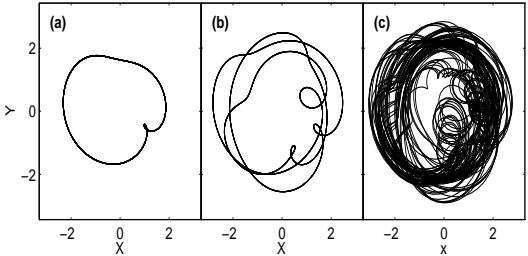,height=3.5cm,width=7cm}
\caption{%
The phase reversing orbit can bifurcate into
a period-3 orbit leading to a chaotic attractor. These
typical states are shown for
(a) $K_2 = 2.2$, (b) $K_2 = 2.5$, (c) $K_2 = 2.712$.
The other constants are
$K_1 = 4.2$, $a = 1$, $\omega = 1$, and $\tau = 0.5$.
}
\label{FIG:prb}
\end{figure}
\begin{figure}
\PSFIG{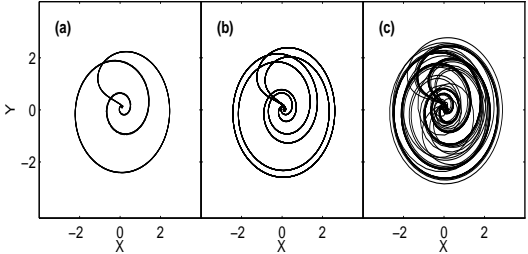,height=3.5cm,width=7cm}
\caption{%
The spiraling orbit period doubles once and then goes to a chaotic
state as typically shown in the sequence,
(a) $K_2 = 3.0$, (b) $K_2 = 3.121$, (c) $K_2 = 3.128$.
The other constants are $K_1 = 0.2$, $a = 1$, $\omega
= 1$, and $\tau = 0.5$.
}
\label{FIG:spb}
\end{figure}
\begin{figure}
\centerline{\psfig{file=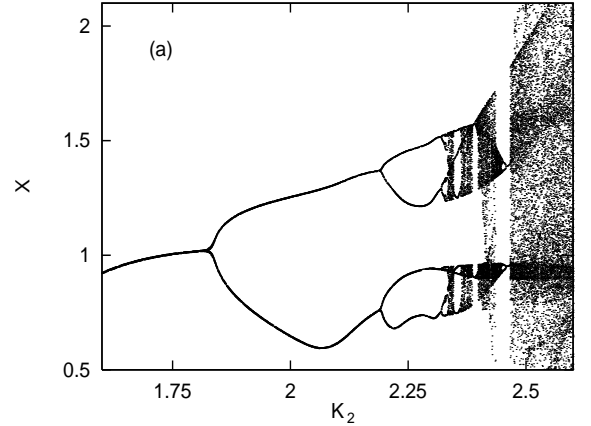,height=6cm,width=8cm,angle=270}}
\centerline{\psfig{file=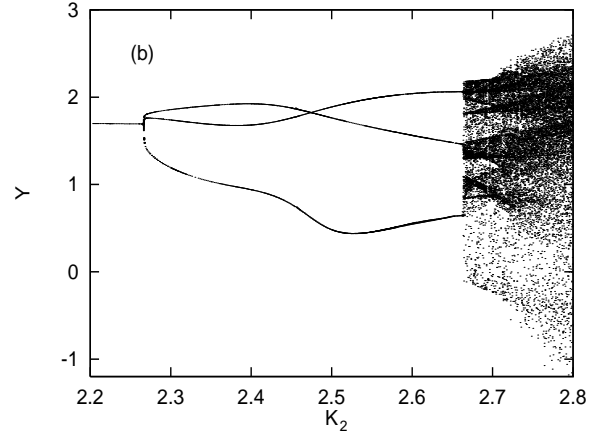,height=6cm,width=8cm,angle=270}}
\centerline{\psfig{file=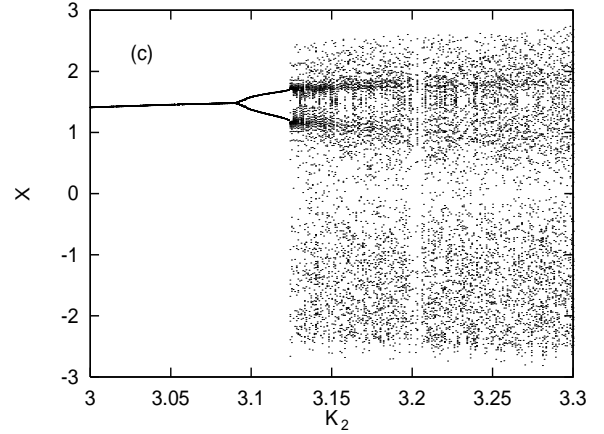,height=6cm,width=8cm,angle=270}}
\caption{%
Typical bifurcation diagrams generated by Poincare section techniques 
showing
(a) phase trapped, (b) phase reversing, and (c) spiraling solutions.
The plots show the value of
(a) $X$ in the plane $Y=0$, $\dot{Y} < 0$, $K_1=-0.25$,
(b) $Y$ in the plane $X=-1.4$, $\dot{X} < 0$, $K_1=4.3$ and
(c) $X$ in the plane $Y=-2.$, $\dot{Y} < 0$, $K_1 = 0.2$.
The other constants are $a = 1, ~\omega = 1,$ and $\tau = 0.5$.
}
\label{FIG:ptprspbd}
\end{figure}
\EM
\begin{figure}
\PSFIG{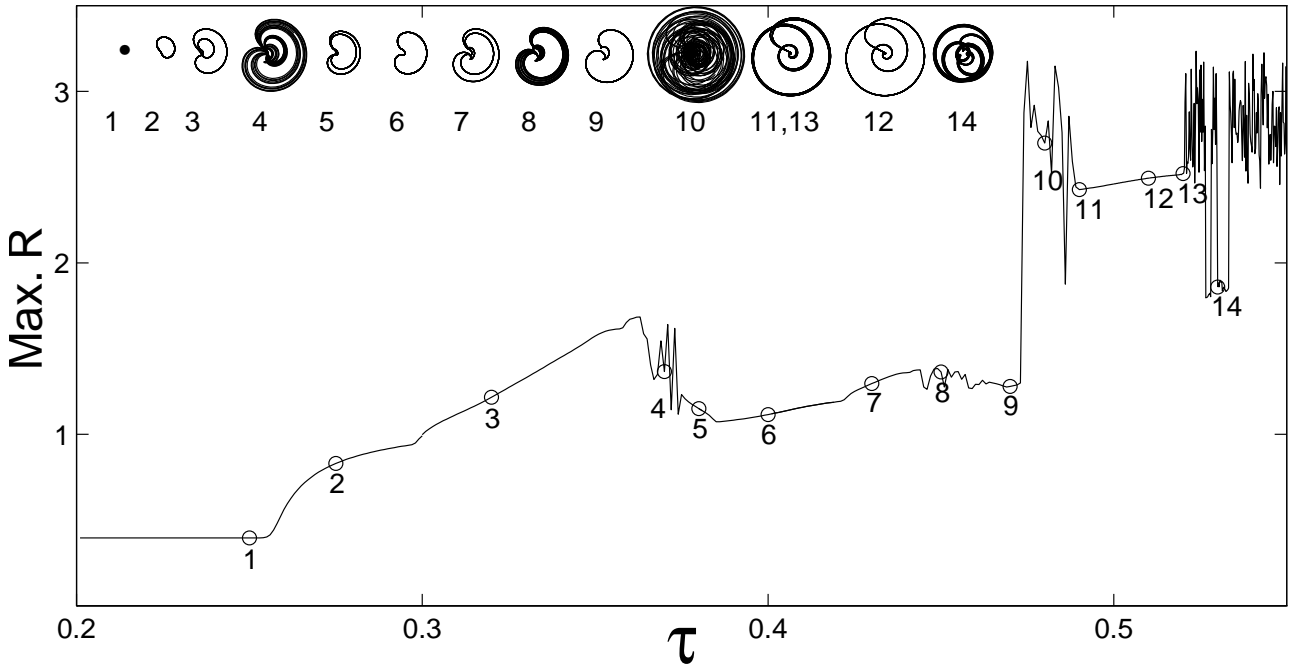,width=18cm} 
\caption{%
Bifurcation diagram as a function of $\tau$.
$R$ is the maximum value of the amplitude of the oscillator
as a function of $\tau$ for $K_2 = -3.0$, and $K_1 = 0.2, a = 1$, and 
$\omega = 1$.
The initial conditions used are $X(0) = 0.123$, $Y(0) = 0.456$.
The orbits at various points that are numbered are shown in the inset.
The orbit at $11$ and $13$ is the spiraling orbit with period-2.
}
\label{FIG:maxRvsTAU}
\end{figure}
\BM
\subsubsection{Routes to Chaos}
The nonlinear model equation (\ref{EQN:E1non}) also exhibits
regions of temporal chaos and we have investigated this
phenomenon in some detail, particularly with regard to routes to
chaos. The route to chaos appears to be a strong function of the
parameters $K_1, K_2$ and $\tau$ and the type of periodic orbit
existing in a particular parameter domain. There is evidence of
several distinct routes to chaos in the present model system. 
For example, the {\it radially trapped}
solutions appear to follow the period
doubling route to chaos as shown in the sequence of plots
in Fig.~\ref{FIG:ptb}. The phase reversing solutions appear to go to
the chaotic state in several different ways. The limit cycle can
make a transition from a simple orbit to a quasi-periodic state.
It can also period double.
Or a period 1 state can make a transition to a period-3 orbit.
Such a transition is shown in Fig.~\ref{FIG:prb}.  We find a large 
window of Period-3 orbits starting from $K_2 = 2.283$ for the 
parameters given in Fig.~\ref{FIG:prb}. The spiraling orbits appear 
to undergo a single period doubling bifurcation and then a sudden 
transition to chaos. An example is shown in Fig.~\ref{FIG:spb}. 
Finally, in Fig.~\ref{FIG:ptprspbd}, we display
the detailed bifurcation diagrams (obtained from Poincare section 
plots) corresponding to these scenarios. 
We would like to add a word here about the numerical precautions 
followed in generating these diagrams. 
Since many of these attractors can coexist in a birhythmic state, it
is necessary to adopt high accuracy and care in tracking any one of 
them.  Likewise one also needs to get beyond some long transients that 
the system can exhibit for some particular initial conditions.
Using these precautions, in Fig.~\ref{FIG:maxRvsTAU}, we have given 
a detailed bifurcation diagram as a function of the time delay 
parameter. In this, the maximum amplitude of the oscillator is 
plotted for $K_1 = 0.2, K_2 = -3.0$ at various values of $\tau$.
The various numbers on the curves stand for the kind of orbit that 
is found in that region (a phase portrait of the orbit is also 
depicted for reference).  The initial conditions chosen have been 
$X(0)= 0.123$ and $Y(0) = 0.456$.
\section{Summary and Discussion}
\label{sec:con}
We have studied the dynamics of a single Hopf bifurcation oscillator
(the Stuart-Landau equation) in the presence of an autonomous 
time delayed feedback. The feedback
term has both a linear component and a simple quadratic nonlinear 
term.  Using a combination of analytical methods and numerical 
analysis, we have
investigated the temporal dynamics of this system in various regimes
characterized by the natural parameters of the oscillator (e.g. its
frequency $\omega$, linear growth rate $a$), strengths of the feedback
components ($K_1$, $K_2$) and the time delay parameter, $\tau$. Our 
principle results are presented in the form of bifurcation diagrams
in these parameter spaces. These reveal a rich variety of temporal
behavior including time delay induced stabilization of the
origin, multiple frequency states, frequency suppression, phase
slips, saddle node bifurcations, and chaotic behavior. In addition,
some of the periodic orbits exhibit novel behavior such 
as birhythmicity, phase reversals, radial trapping, 
spiraling oscillations in amplitude
space. 
Some of these can be understood by the stability of the fixed points
or the loss of single valuedness of amplitude evolution equation.

One of the attractive features of these results is that many of them
have been observed in the  collective behavior of larger systems such
as the Kuramoto model or the amplitude versions of the Kuramoto model. 
This has been one of our major motivations for constructing this
model - as a sort of paradigm to obtain and investigate these
states in a simple manner.  The feedback terms not only model the
collective drive that a single oscillator feels in a larger system,
but also incorporate time delay in an autonomous manner. Time delay
frees the dimensional constraints of the system (the system is 
essentially $\infty$ -dimensional) and this might be the reason why
its temporal dynamics resembles so much that of larger
dimensional systems. Our results may therefore be useful for gaining
better insights into the behavior of such large systems. As an 
example, the large window of $W-3$ orbits of our model appears to 
have a correspondence
to the transition region between the incoherent and chaotic states of
coupled limit cycle oscillators. We have found that the oscillators
in the wings of the frequency distribution of a large collection
of oscillators in the mean field model begin to
acquire the $W-3$ temporal states in that region and play an
important role in the transition mechanism \cite{RSJ:new}.  A
detailed understanding of their dynamical behavior can thus
help us in addressing some of the outstanding problems in this
area, such as the nature of transition between low dimensional
chaos and turbulence. Our model can also find more direct
applications in simulation studies for feedback control of
individual physical, chemical or biological entities that have the
basic nonlinear characteristics of our Hopf oscillator,
such as in single mode semiconductor lasers, relativistic magnetrons,
chemical oscillations, and biological rhythms in single nerve cells.
In fact, the basic Stuart-Landau equation 
is mathematically related to the well known van der Pol
oscillator equation from which it can be derived by a suitable 
time averaging.  Our oscillator model with the quadratic nonlinearity 
can likewise be derived from
a van der Pol type equation which in addition has
a nonlinear Mathieu like term, i.e., it is an equation of the form
\be
\ddot{x}+\mu ~(x^2-1) \dot{x}+(1+~\epsilon ~\dot{x} ~\sin(t) ) ~x = 0 .
\ee
Such a nonlinear equation can physically
represent parametric excitation of relaxation oscillations and can
be used to model a number of physical or biological systems. It
may be possible in such simple systems to then seek experimental
verification of some of the novel temporal states displayed by
our model.

\end{multicols}
\end{document}